\documentclass[a4paper]{a2}
\usepackage{graphicx}  
\begin{document}
\def\ea{et~al.}
\def\gi{\; \mathrm{Gyr}}
\newcommand{\rbox}[1]{\raisebox{1.5ex}[0pt]{#1}}


\title{PG 1613+426: a new sdB pulsator\thanks{Based on 
observations carried out at the Catania Astrophysical Observatory {\it M. G. Fracastoro}
stellar station on Mt. Etna, Italy}}
   \subtitle{}
   \author{A.~Bonanno\inst{1} \and S.~Catalano \inst{1} 
   \and A.~Frasca\inst{1} \and G.~Mignemi\inst{2} \and L.~Patern\`o\inst{2} }

\offprints{A.~Bonanno, \email{abo@ct.astro.it}}
\institute{INAF - Osservatorio Astrofisico di Catania, Citt\`a Universitaria, 
I-95123 Catania, Italy \and Dipartimento di Fisica e Astronomia 
dell'Universit\`a, Sezione Astrofisica, Citt\`a Universitaria, I-95123 Catania, Italy}
\date{\today}

\authorrunning{A.~Bonanno et al.}
\titlerunning{PG 1613+426: a new sdB pulsator}

\abstract{We report the detection of short period oscillations in the hot subdwarf B (sdB) star 
PG 1613+426 from  time-series photometry carried out with the 91-cm 
Cassegrain telescope of the Catania Astrophysical Observatory. 
This star, which is brighter than the average of the presently known sdB pulsators, 
with B = 14.14 mag, has $T_{\rm eff}=34\,400\,{\rm K}$ and 
$\log g = 5.97$, its position is near the hot end of the 
sdB instability strip, and it
is a pulsator with a well observed peak in
the power spectrum at $144.18\pm 0.06\,\rm s$. This star seems to be well suited for high precision measurements, which could detect a possible multi-mode pulsation behaviour. 

}
\maketitle

\section{Introduction}
The hot subdwarf B (sdB) stars form 
a homogeneous group populating an extension of the extreme horizontal
branch (EHB) in the $(T_{\rm eff}- \log g)$-diagram towards temperatures
up to $40\,000\,{\rm K}$.  
These stars are evolved low-mass $(\sim 0.5 M_\odot)$ objects with a He burning 
core surrounded by a H surface layer which is too thin ($< 4\%$ by mass)
to sustain the H-shell burning. Their origin is still debated, but it seems
that they have experienced He-flash phase and a substantial mass loss along the 
red giant branch. Their further evolution proceeds toward the EHB by crossing
the subdwarf O population and eventually entering the white-dwarf graveyard
(Maxted et al. 2001, Heber et al. 2002).

The recent discovery that several of them are multimode pulsators has triggered 
both observational and theoretical efforts for studying their characteristics
and pulsating mechanisms (Charpinet et. al. 2001).  
Data on 30 of such pulsators, also known as
sdB variables (sdBVs), are reported by Charpinet (2001), including 
the sdBV HS0702+6043, the detailed results of which are reported in Dreizler et al. (2002), and the two sdBVs PG1325+101 and PG2303+019, recently discovered by Silvotti et al. (2002). 
The 31st star of this class was discovered by Piccioni et al. (2000) and later confirmed by Ulla et al. (2001).
These stars show short period ($1-10\,\rm min$) and low amplitude (1-50 milli-magnitude [mma]) non-radial pulsation modes. Charpinet et al. (1996,1997) have shown that the non-radial pulsations are probably driven by an opacity bump associated 
with iron ionization.

Here we report the discovery of a new variable of this class, the 32nd one,
namely the PG~1613+426 sdB, thanks to the data collected 
at the Catania Astrophysical Observatory. This star, with B = 14.14 mag, is brighter than the average of the presently known sdB pulsators. 
This object has been selected from the list of ultraviolet-excess 
Palomar Green objects (Saffer et. al 1994), where
a determination of the main atmospheric parameters is available. 
For this star, low resolution spectroscopy ($\sim$~6\,\AA) leads to a sdOA spectral classification 
due to its strong, broad Balmer and HeI absorption lines, and in particular gives
$T_{\rm eff}=34\,400\,\pm\,500\,{\rm K}$, $\log g = 5.97 \pm 0.12$, and
$\rm N(He)/N(H)=0.022\,\pm\,0.003$ from a least-square fitting of Balmer lines
with LTE models. These characteristics place this objects near the hot 
end of the theoretical sdBVs instability strip.  

\section{Photometric observations}
The photometric observations were carried out with the the 91-cm Cassegrain telescope 
of the {\it M. G. Fracastoro} stellar station  
of the Catania Astrophysical Observatory (Serra la Nave, Mt. Etna, $1720\,\rm m$ a.s.l.).
The telescope was equipped with a photon counting photometer, which used an 
unfiltered EMI 9789-QA photomultiplier as detector. A 21$\arcsec$ diaphragm  was used to isolate the star light from the sky background.
Owing to the PG~1613+426 spectral distribution and the response curve of the photocathode,
the photometric passband included from B to UV  wavelengths. 
Data were collected by integrating in time intervals of 15 or $20\,\rm s$, obtaining typical
counts of about $20\,000$ per measurement.
The estimated photometric accuracy at our observing station 
was approximately $5$ mma in good observing nights.
Exposures of a reference star in the same stellar field and sky background were also taken during the night in order to subtract the sky background and check the overall stability of the photometric system.
The correction for atmospheric extinction was applied by using the average seasonal extinction coefficient for B Johnson filter. Since our spectral band did not 
coincide with the Johnson B band and the
variation of the sky transparency was not fully taken into account by the non-simultaneous measurements of the 
comparison star, possible long-term trends of differential magnitude
were removed by subtracting a low-order polynomial curve fitted to the data 
of the individual runs. 

\section{Data analysis and results}
\begin{figure}[t]
\resizebox{\hsize}{!}{\includegraphics{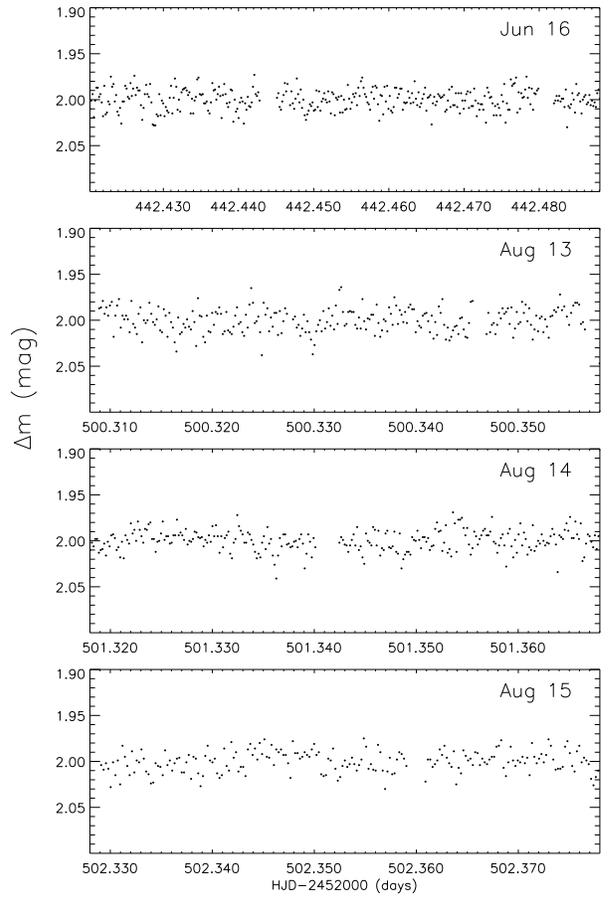}}
\caption{Sample of the differential light curves of PG~1613+426. Long term trends
have been subtracted with low order polynomial fitting.}
\label{fig:cl}
\end{figure}

PG~1613+426 was identified as a variable star on 2002 June 14.
We have thus decided to observe again this target on June 15 and June 16. Further observing 
runs were then performed in August 8, 9, 13, 14 and 15 in order to resolve the amplitude spectra with a better accuracy. A summary of the observations, including the starting time, date, duration, exposure time per measurement, name of the observers and the main pulsation period, is reported in Table 1, while a sample of the differential light curves is shown in Fig.~\ref{fig:cl}.

\begin{table}
\caption{Summary of the observations. The initials of the observers reported below are: AB = A. Bonanno, AF = A. Frasca, GM = G. Mignemi.}
\label{tab:log}
\begin{center}
\begin{tabular}{clrclc}
\hline
\noalign{\medskip}
{HJD$_{\rm start}$ } &  {Date}  &  {Length} & ${T_{\rm exp}}$ &   
{Observers} & $P_{\rm max}$ \\
  -2452000          & (2002)       &    (s)~~   &   (s)     &               &  (s) \\
\hline
\noalign{\medskip}
  440.53128	& Jun 14  &   1940   &     15	 &	GM	 & 143.7$\,\pm\,$5.7 \\
  441.51470	& Jun 15  &   2762   &     15	 &	 GM	 & 142.3$\,\pm\,$3.8 \\
  442.42023	& Jun 16  &   5861   &     15	 &	GM	 & 144.0$\,\pm\,$1.8 \\
  495.33374	& Aug 08  &   2424   &     15	 &    GM\,+\,AF	 & 144.8$\,\pm\,$4.6 \\
  496.34967	& Aug 09  &   3937   &     15	 &	 GM	 & 143.9$\,\pm\,$1.6 \\
  500.30893	& Aug 13  &   11804  &     15	 &    GM\,+\,AB	 & 144.5$\,\pm\,$0.9 \\
  501.31765	& Aug 14  &   4655   &     15	 &    GM\,+\,AB	 & 144.2$\,\pm\,$1.0 \\
  501.37527	& Aug 14  &   5426   &     20	 &    GM\,+\,AB	 & 144.2$\,\pm\,$1.0 \\
  502.32910	& Aug 15  &   9392   &     20	 &    GM\,+\,AB	 & 144.3$\,\pm\,$1.2 \\
\noalign{\medskip}
\hline
\end{tabular}
\end{center}
\end{table}

\begin{figure}[t]
\resizebox{\hsize}{!}{\includegraphics{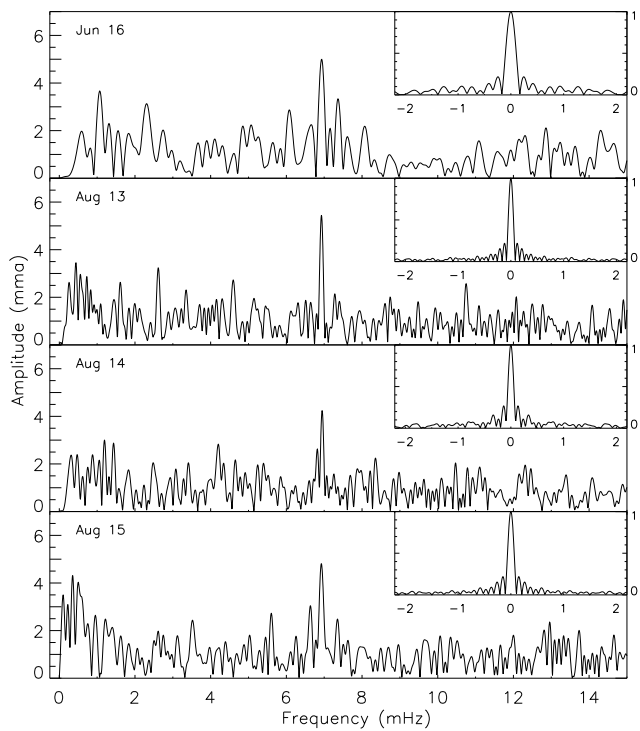}}
\caption{Amplitude spectra for some observing nights. The spectral window is shown 
in the insets.}
\label{fig:dft}
\end{figure}

\begin{figure}[t]
\resizebox{\hsize}{!}{\includegraphics{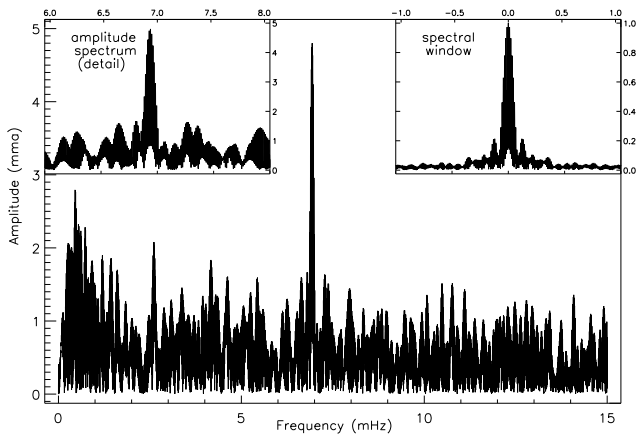}}
\caption{Amplitude spectrum for the three August consecutive nights. The spectral window and a
detail of the spectrum around its maximum are shown in the insets.}
\label{fig:dft_tot}
\end{figure}

The results concerning the amplitude spectra are illustrated in Fig.~\ref{fig:dft} for 
the three individual August 13, 14 and 15 nights and the June 16 night. The amplitude spectrum of the
three August consecutive nights, during which the observing conditions were particularly good, taken toghether, is shown in Fig.~\ref{fig:dft_tot}. The amplitude spectrum  
shows a clear, dominant peak at $6.936\pm 0.003\,{\rm mHz}$ ($144.18\pm 0.06\,\rm s$) with an amplitude of 
about 5 mma in each observation night. The period corresponding to the highest peak in
the power spectrum is reported in Table 1 for each night. The error has been estimated from the FWHM of the main peak in the spectral window. The Lomb-Scargle periodogram analysis 
(Scargle \cite{Sca82}) applied to our data gives 
a confidence level $\geq 99.9999\,\%$ for the highest peak in 
the spectrum obtained in the June 16, and August 13, 14 and 15 observation runs.

We have refined our analysis by pre-withening the light curves
with a sinusoidal fit of the dominant period, and iterating to obtain the light curve residuals for any new frequency which was found
\footnote{We used the software Period98 by M. Sperl, University of Vienna, 1998.}.
We found additional frequencies, but we noticed that the associated reduced $\chi^2$ did not change significantly during the pre-whitening procedure, indicating that, at the present level of signal to noise ratio, the analysis precludes any further peak identification. 

As a complementary method we used the CLEAN iterative deconvolution algorithm 
(Roberts et al. \cite{Rob86}) in order to eliminate the effects of the observational 
spectral window in the  power spectrum. The cleaned periodograms of the individual
nights do not differ significantly from the non-cleaned spectra, since the spectral
windows for essentially equally spaced data are always rather clean with
side-lobes of low amplitude. The results of this analysis, applied to the three August consecutive nights, do not reveal any difference
from the previous one, confirming the clear presence of the peak at $144.18\,\rm mHz$.

Although the accuracy of our measurements
is too low to reach a firm conclusion on the existence of additional periods, we cannot rule out the possibility that PG~1613+426 is a multimode pulsator. For instance, we note that 
frequencies in the range  $6.63-7.92\,\rm mHz$   
are expected on the basis of the theoretical models discussed 
in Charpinet et al. (2001) and on observations of 
very similar spectra in other hot sdBVs
which have nearly the same position  
in the sdBVs instability strip, like HS~0815+4243, with
$T_{\rm eff}=33\,700\,\rm K$ and $\log g = 5.95$, and HS~2149+0847, with
$T_{\rm eff}=35\,600\,\rm K$ and $\log g = 5.9$ (\O stensen et al. 2001).

We think that it would be important to observe this object by means of high precision 
CCD photometry in order to resolve its spectrum with better accuracy, and  
we plan to pursue this project with observations at our site in the near future.

\acknowledgement{We gratefully acknowledge the referee, U. Heber, for useful suggestions concerning the revision of the manuscript and R. Silvotti for many enlightening discussions.
We thank the Italian
{\em Ministero dell' Istruzione, Universit\`a e Ricerca} and the {\em Regione Sicilia} for their support.
This research made use of the SIMBAD and VIZIER databases, operated at CDS, 
Strasbourg, France.}


\begin{thebibliography}{}

\bibitem[2001]{car2} Charpinet, S. 2001, Astron. Nachr., 322, 387

\bibitem[2001]{car}  Charpinet, S., Fontaine, G., \& Brassard, P. 2001, PASP, 113, 775

\bibitem[2001]{car3} Charpinet, S., Fontaine, G., Brassard, P., \& Dorman, B. 1996, ApJ 471, 	L106

\bibitem[2001]{car4} Charpinet, S., Fontaine, G., Brassard, P., et.al. 1997, ApJ, 483, L123

\bibitem[2002]{} Dreizler, S., Schuh, S. L., Deetjen, J. L., Edelmann, H., \& Heber, U. 2002, 	A\&A, 386, 

\bibitem[2002]{} Heber, U., Moehler, S., Napiwotzki, R., Thejll, P., \& Green, E. M. 2002, 	A\&A,	383, 938

\bibitem[2001]{} Maxted, P. F. L., Heber, U., , Marsh, T. R., \& North, R. C. 2001, MNRAS, 326, 	1391

\bibitem[2001]{oe}  \O stensen, R., Solheim, J.-E., Heber, U., et al. 2001, A\&A, 368, 175

\bibitem[2000]{} Piccioni, A., Bartolini, C., Bernabei, S., et al. 2000, A\&A, 354, L13

\bibitem[1986]{Rob86}Roberts, D. H., Leh\'ar J., \& Dreher J. W. 1986, AJ, 93, 968

\bibitem[1996]{sa}   Saffer, R. A., Bergeron, P., Koester D., \& Liebert, J. 1994, ApJ, 432 351

\bibitem[1982]{Sca82}Scargle, J. D. 1982, ApJ, 263, 835

\bibitem[2002]{sil} Silvotti, R. \O estnsen, R., Heber, U., et al. 2002, A\&A, 383, 239

\bibitem[2001]{} Ulla, A., Zapatero Osorio, M. R., Hern\'andez P\'erez, F., MacDonald, J. 2001, 	A\&A, 369, 986

\end{thebibliography}
\end{document}